\def\year{2022}\relax
\documentclass[letterpaper]{article} 
\usepackage{aaai22}  
\usepackage{times}  
\usepackage{helvet}  
\usepackage{courier}  
\usepackage[hyphens]{url}  
\usepackage{graphicx} 
\usepackage{natbib}  
\usepackage{caption} 
\DeclareCaptionStyle{ruled}{labelfont=normalfont,labelsep=colon,strut=off} 
\usepackage{algorithm}
\usepackage{algpseudocode}

\usepackage{subcaption}
\usepackage{bm}

\usepackage{newfloat}
\usepackage{listings}
\floatstyle{ruled}
\newfloat{listing}{tb}{lst}{}
\floatname{listing}{Listing}
\title{ML-based Anomaly Detection in Optical Fiber Monitoring}
\author{
    Khouloud Abdelli \textsuperscript{\rm 1,2}, 
    Joo Yeon Cho \textsuperscript{\rm 2},
    Carsten Tropschug \textsuperscript{\rm 3}
}
\affiliations{
    \textsuperscript{\rm 1} Christian-Albrechts-Universität zu Kiel, 
Kaiserstr. 2, Kiel, Germany, 24143 \\

\textsuperscript{\rm 2} ADVA Optical Networking SE, 
Fraunhoferstrasse 9a, 
Martinsried, Germany, 82152 \\

\textsuperscript{\rm 3} ADVA Optical Networking SE,
Märzenquelle 1-3, 
Meiningen, Germany, 98617 \\
\{KAbdelli, JCho, CTropschug\}@adva.com 
}

\usepackage{bibentry}

\begin{document}

\maketitle

\begin{abstract} 
Secure and reliable data communication in optical networks is critical for high-speed internet.
We propose a data driven approach for the anomaly detection  and faults identification  in optical networks to diagnose
physical attacks such as fiber breaks and optical tapping.
The proposed methods include  an autoencoder-based anomaly detection and an attention-based  bidirectional gated recurrent unit algorithm for 
the fiber fault identification and localization.
We verify the efficiency of our methods by experiments under various attack scenarios using real operational data. 
 
\end{abstract}

\section{Introduction} \label{sec_intro}
Optical fiber is the essential medium for transporting a large amount of data through the aggregated internet, mobile backhaul and core network.
A single fiber link connects thousands of customers and enterprises, carrying a mixture of personal, business and public data. Therefore,
the impact of a broken fiber can be enormous and must be responded to immediately. 

It is well known that an optical fiber is vulnerable to various types of attacks such as fiber cut and fiber tapping (eavesdropping).
Such attacks 
compromise the availability and the confidentiality of an optical network.
Specifically, a manual discovery of incidents occurring in the fiber require considerable expert knowledge and probing time until 
a fault (e.g. broken fiber) is identified.

Fiber monitoring aims at detecting anomalies in an optical layer by logging and analyzing the monitoring  data. 
It has mainly been performed using optical time domain reflectometry (OTDR), a technique based on Rayleigh backscattering, widely applied for fiber characteristics’ measurements and for fiber fault detection and localization \cite{Lee:14}. 
OTDR operates like an optical radar. 
It sends a series of optical pulses into the fiber under the test. 
The backscattered signals are then recorded as a function of time, which can be converted into the position on the optical fiber. 
As result, a recorded OTDR trace illustrating the positions of faults along the fiber, is generated, and used for event analysis. 

However, OTDR traces are difficult to interpret even by highly experienced engineers mainly due to the noise overwhelming the signals. Analyzing OTDR signals using conventional methods can be time consuming as performing a lot of averaging of OTDR measurements is required to remove the noise and thereby to achieve a good event detection and localization accuracy. 
Therefore, it would be highly beneficial to develop a reliable automated diagnostic method that accurately and quickly detects, diagnoses, and pinpoints fiber faults given the OTDR data. 
It reduces operation-and-maintenance expenses (OPEX) and eliminates
the time needed to investigate the cause and determine a search area.
Upon finding the fault location, appropriate action is taken to remedy the fault and restore service as quickly as possible.

Recently, machine learning (ML) based approaches have shown great potential to tackle the problem of fiber event detection and localization \cite{nyarko2021predicting}. In this respect, long short-term memory and convolutional neural networks have been proposed to detect and localize the reflective fiber events induced by the connectors and mechanical splices \cite{abdelli2021reflective,abdelli2021conv}.  

\begin{figure}[!t]
\centering
\includegraphics[width=1.0\columnwidth]{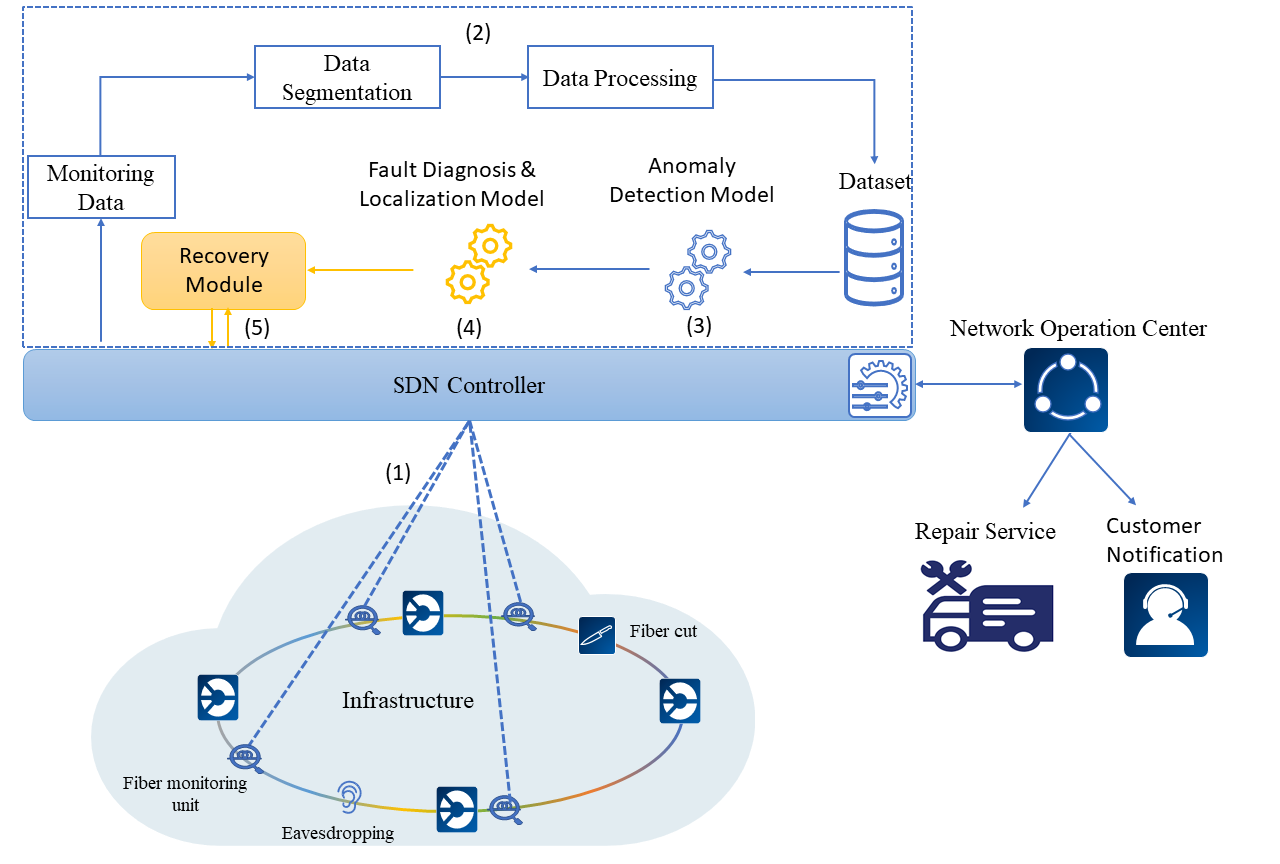}
\caption{Overview of the ML-based fiber monitoring process}
\label{fig-1}
\end{figure}

In this paper, we propose an ML-based fiber fault detection, localization, and diagnosis framework, leveraging OTDR data.
The proposed approach includes an autoencoder to detect 
a fiber fault (e.g. fiber cut or bending) 
and an attention-based bidirectional gated recurrent unit model to identify the type of the detected fault and localize it. 
The proposed framework is validated by noisy experimental OTDR data with the SNR varying from 0 dB to 30 dB.
Fig. \ref{fig-1} shows that an overview of the ML-based fiber monitoring process.

\section{Background} \label{sec-bg}
\paragraph{Multi-Task Learning} 
Multi-task learning (MTL) is a learning paradigm in ML that aims to improve the generalization performance of multiple tasks by learning them jointly while sharing knowledge across them. 
It has been widely used in various fields ranging from natural language processing to computer vision. 
The MTL approaches can be classified as hard- and soft parameter sharing. The hard parameter sharing method is done by sharing the hidden layers with the different tasks (completely sharing the weights and the parameters between all tasks) while preserving task-specific output layers learnt independently by each task. Whereas for the soft parameter sharing approach, a model with its own parameters is learned for each task and the distance between the parameters of the model is then regularized to encourage similarities among related parameters \cite{ruder2017overview}. 

\paragraph{Autoencoder}
An autoencoder (AE) is a type of artificial neural network seeking to learn a compressed representation of an input in an unsupervised manner \citep{https://doi.org/10.1002/aic.690370209}. 
An AE is composed of two sub-models namely the encoder and the decoder. The encoder is used to compress an input $X$ into lower-dimensional encoding (i.e. latent-space representation) $Z$ through a non-linear transformation, which is expressed as follows:
\begin{equation}
\bm{Z}=f( \bm{WX} + \bm{b}),
\end{equation}
where $\bm W$ and $\bm b$ denote the weight and bias matrices of the encoder and $f$ represents the activation function of the encoder.

The decoder attempts to reconstruct the output $\hat{X}$ given the representation $\bm Z$ via a nonlinear transformation, which it is formulated as follows:
\begin{equation}
\hat{\bm X} = g( \bm{W'X} + \bm{b'} ),
\end{equation}
where $\bm W'$ and $\bm b'$ represent the weight and the bias matrices of the decoder and $g$ denotes the activation function of the decoder.

The AE is trained by minimizing the reconstruction error between the output $\hat{bm X}$ and the input $\bm X$, which is the loss function $L(\theta)$, typically the mean square error (MSE), defined as:
\begin{equation}
L(\theta)= \sum || \bm{X} - \hat{\bm X} ||^2    
\end{equation}
where $\theta = \{\bm{W},\bm{b},\bm{W'},\bm{b'}\}$  denotes the set of the parameters to be optimized.

\paragraph{Bidirectional GRU}
Gated recurrent unit (GRU) is a specific type of Recurrent Neural Networks (RNNs) to solve the problem of gradient vanishing \citep{cho2014learning}. 
It has been widely adopted to process sequential data and to capture long-term dependencies. 
The typical structure of GRU is composed of two gates namely reset and update gates, controlling the flow of the information. 
The update gate regulates the information that flows into the memory, while the reset gate controls the information flowing out the memory. The GRU cell is updated at each time step t by applying the following equations:
\begin{eqnarray}
\bm{z}_t &=& \sigma(\bm{W}_z \cdot x_t + \bm{W}_z \cdot \bm{h}_{(t-1)}  + \bm{b}_z) \\
\bm{r}_t &=& \sigma(\bm{W}_r \cdot \bm{x}_t + \bm{W}_r \cdot \bm{h}_{(t-1)}  + \bm{b}_r) \\
\hat{\bm{h}_t} &=& {tanh}( \bm{W}_h \cdot \bm{x}_t + \bm{W}_h \cdot (\bm{r}_t \circ \bm{h}_{(t-1)}) + \bm{b}_h) \\
\bm{h}_t &=& \bm{z}_t \circ \bm{h}_{(t-1)} + (1-\bm{z}_t) \circ \hat{\bm{h}_t}   
\end{eqnarray}
where $\bm{z}_t$ denotes the update gate, $\bm{r}_t$ represents the reset gate, $\bm{x}_t$ is the input vector, $\bm{h}_t$ is the output vector, $\bm{W}$ and $\bm{b}$ represent the weight and the bias matrices respectively. 
$\sigma$ is the gate activation function and $tanh$ represents the output activation function. 
The '$\cdot$' operator denotes a matrix multiplication and the '$\circ$' operator represents the dot product. 

Bidirectional GRU (BiGRU) is an extension of GRU that helps to improve the performance of the model. 
It consists of two GRUs: one is a forward GRU model that takes the input in a forward direction, and the other is a backward GRU model that learns the reversed input. 
The output $y_t$ of the model is generated by combining the forward output $\overrightarrow{ h_t}$ and the backward output $\overleftarrow{h_t}$ as described by the following equations:
\begin{eqnarray}
\overrightarrow{\bm{h}_t}&=& GRU ( \bm{x}_t, \overrightarrow{\bm{h}_{t-1}} ) \\
\overleftarrow{\bm{h}_t}&=& GRU ( \bm{x}_t, \overleftarrow{\bm{h}_{t-1}}) \\
\bm{y}_t &=& \overrightarrow{\bm{h}_t} \oplus \overleftarrow{\bm{h}_t}
\end{eqnarray} 
where $\oplus$  denotes an element-wise sum.

\section{Physical Attacks on Fiber}
\paragraph{Fiber cut:} Intentional fiber cut attack is an attempt to deny or disrupt service by cutting the fiber optic cables and thereby inducing a widespread denial of service attack. 
Fiber break is considered as the single largest cause of service outages. 
As reported by the Federal Communication Commission (FCC), more than one-third of service disruptions are caused by fiber-cable breaks \cite{2007OptLE..45..126B}. 
Any service outage due to a fiber cut results in massive data loss, network disruption, and huge financial loss etc \cite{759416}. 
In 1991, a severed fiber-optic cable shut down all the New York airports and induced air traffic control problems \cite{gorman2005networks}. 
It is time-consuming to locate and repair fiber breaks. 

\paragraph{Optical eavesdropping:} Optical eavesdropping attack permits the eavesdropper to gain an unauthorized access to the carried data by directly accessing the optical channel via fiber tapping for the purpose of stealing mission-critical and sensitive information. 
There are several fiber tapping techniques which can be adopted to launch the eavesdropping attack, such as fiber bending, optical splitting, evanescent coupling, V Groove cut and so on \citep{1494884}. 
However, the easiest method to make the eavesdropping intrusion undetected is micro-fiber bending by using commercially available clip-on coupler. 
Fiber-optic cable tapping incidents have been reported such as the eavesdropping device, which was discovered illegally installed on Verizon's optical network in 2003 to glean information from a mutual fund company regarding its quarterly statement prior to its public disclosure \cite{hacklight}. 
Although, it is easy to perform an eavesdropping intrusion, it is challenging to detect such intrusion using conventional intrusion detection methods such as 
OTDR-based techniques.

\section{Fiber Monitoring Framework}
As shown in Figure \ref{fig-1}, 
the proposed framework can be broken into five main stages: (1) optical fiber network monitoring and data collection, (2) data processing, (3) fiber anomaly detection, (4) fiber fault diagnosis and localization, (5) mitigation and recovery from fiber failures. 
The optical fibers deployed in the network infrastructure are periodically monitored using OTDR (i.e fiber monitoring unit). 

The generated OTDR traces (i.e monitoring data) are sent to the software-defined networking (SDN) controller managing the optical infrastructure. 
Then, the said data is segmented into fixed length sequences and normalized. 
Afterwards, the processed data is fed to the ML based anomaly detection model for recognizing the fiber faults. 
If a fiber failure is detected, a ML model for fault diagnosis and localization is adopted to identify the fault and pinpoint it. 
Based on the identified failure, a set of recovery rules is applied to mitigate such fault. 
The SDN controller notifies the network operation center in case of failure, which informs the customer about the type of detected fault and its location, and notifies the maintenance and repair service in case of fiber break. 

For this work, we consider the physical fiber attacks namely fiber cut and optical eavesdropping attacks, as examples of harmful fiber faults. Given that the patterns of faults namely bad splice and dirty connector are similar to the physical attacks’ patterns particularly under very low SNR conditions, we include them during the training phase of the ML model for fault diagnosis to ensure a reliable fault identification and reduce the false alarms. 
	\subsubsection{Autoencoder-based Anomaly Detection}
The proposed ML model for fiber anomaly detection is based on autoencoder. 
The autoencoder is trained with only normal data representing the normal behavior in order to learn the distribution characterizing the normal state. 
After the training of the autoencoder and for the inference phase, the reconstruction error is adopted as an anomaly score to detect any potential fault. 
A well-trained autoencoder will reconstruct the normal instances very well since they will have the same pattern or distribution as the training data, while it will fail to reproduce the anomalous observations by yielding high reconstruction errors. 
The process of the classification of an instance as anomalous/normal is illustrated in Algorithm 1. 
If the computed anomaly score is higher than a set threshold $\theta$, the instance is classified as "anomalous", else it is assigned as "normal". 
$\theta$ is a hyperparameter optimized to ensure high detection accuracy and is adjusted by taking into consideration the degradation and the aging effect of the optical fiber. 
\begin{algorithm}
\caption{Autoencoder-based anomaly detection}\label{alg:ae}
\begin{algorithmic}
\State {\bf Input }  
Normal dataset $\bm{x}$, Anomalous dataset $\bm{x}^{(i)}, i=1,\ldots, N$, threshold $\theta$
\State {\bf Output} 
Reconstruction error $||\bm{x}-\hat{\bm{x}}||$
\State
$f, g \gets$ train an autoencoder using the normal dataset $\bm{x}$
\For {$ i = 1,  \cdots, N$}
	\State reconstruction error(i) $\gets$ $|| \bm{x}^{(i)} - g \circ f(\bm{x}^{(i)}) || $
	\If {reconstruction error(i) $> \theta$}
		\State $\bm{x}^{(i)}$ is anomalous.
	\Else
		\State $\bm{x}^{(i)}$ is normal.
	\EndIf
\EndFor
\end{algorithmic}
\end{algorithm}

\begin{figure}[!t]
\centering
\includegraphics[width=1.0\columnwidth]{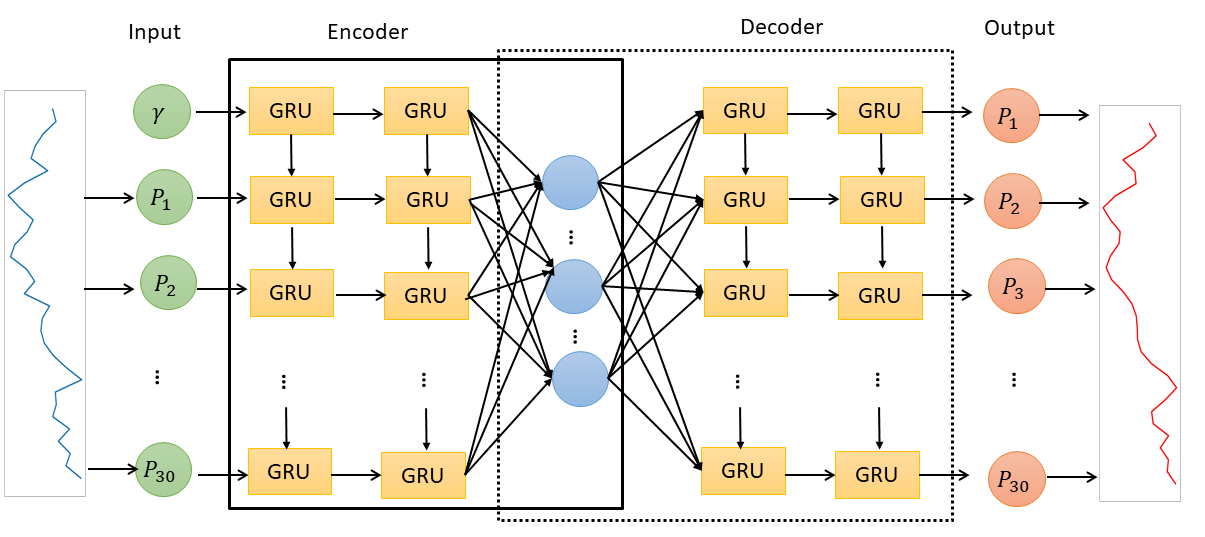}
\caption{Structure of the proposed gated recurrent unit based autoencoder for optical fiber anomaly detection.}
\label{fig-2}
\end{figure}

The architecture of the proposed autoencoder model for fiber anomaly detection is illustrated in Figure \ref{fig-2}. 
The model contains an encoder and a decoder sub-model with 4 layers. 
The encoder takes a sequence of OTDR traces $[ P_1, P_2,\ldots, P_{30}]$ as an input, representing the attenuation of the fiber along the distance, combined with the sequence’s computed 
SNR ($\gamma$). 
The information about the sequence’s SNR during the training phase helps the ML model to learn the behavior of the normal signal pattern for each input SNR level and thereby to boost the performance \cite{abdelli2021reflective}. 
The fed input to the encoder is then compressed into a low dimensional representation by adopting 2 GRU layers composed of 64 and 32 cells, respectively, which captures the relevant sequential features modeling the normal state under different SNRs. 
Afterwards, the decoder reconstructs the output, given the compressed representation output of the encoder. 
The decoder is inversely symmetric to the encoder part. 
Rectified Linear Unit (ReLU) is selected as an activation function for the hidden layers of the model. 
The cost function is set to the mean square error (MSE), which is adjusted by the Adam optimizer.

	\subsubsection{Fault Diagnosis and Localization}
Let us denote the fault diagnosis task by $T_1$ and the fault position estimation task by $T_2$.
Obviously, $T_1$ and $T_2$ can get benefits each other by sharing their features.
The proposed model for $T_1$ and $T_2$ is an MTL framework which 
can learn these tasks simultaneously, enhancing their generalization capability.  
The architecture of the proposed framework is composed of the shared hidden layers distributing the knowledge across $T_1$ and $T_2$ followed by a specific task layer. 
The shared hidden layers 
adopt a combination of a bidirectional gated recurrent unit (BiGRU) network and attention mechanisms. 
The BiGRU is used to capture the sequential features characterizing the pattern of each fault, whereas the attention mechanisms help the model to concentrate more on the relevant features to improve the fault diagnosis and localization accuracy. 

\begin{figure}[!t]
\centering
\includegraphics[width=1.0\columnwidth]{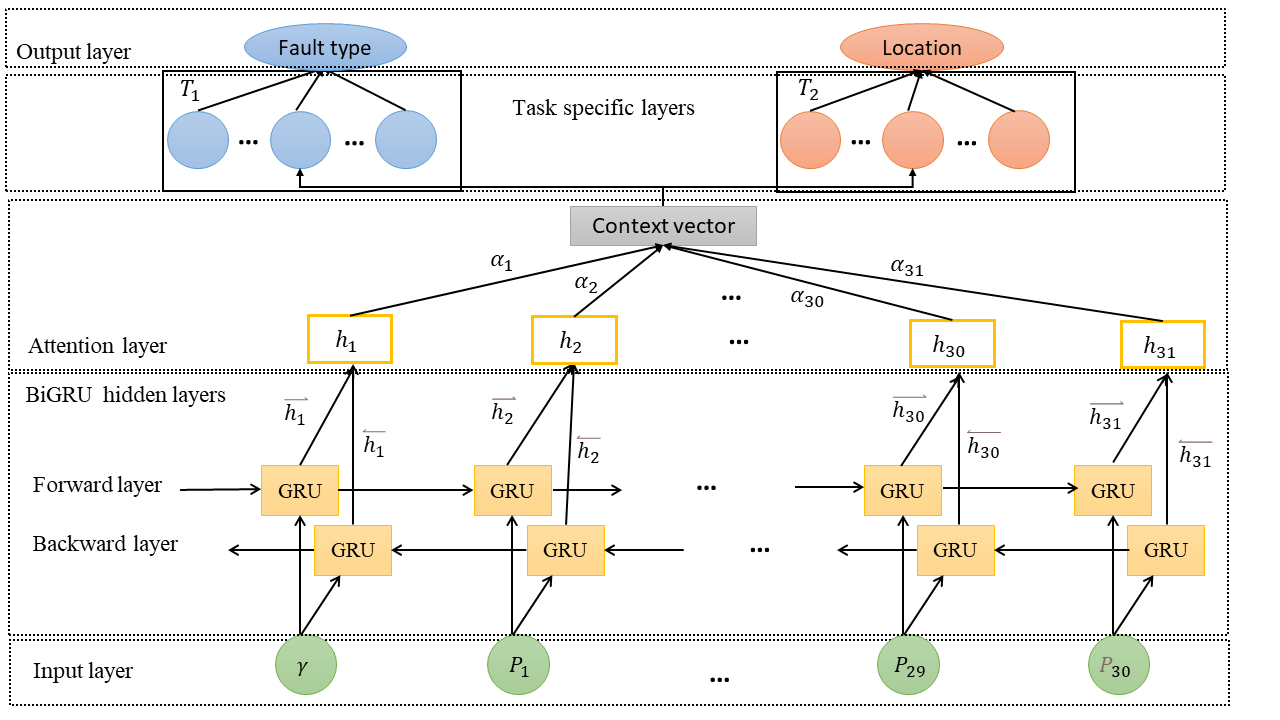}
\caption{Structure of the proposed attention-based bidirectional gated recurrent unit model for fiber fault diagnosis and localization}
\label{fig-3}
\end{figure}

As shown in Figure \ref{fig-3}, 
the input (i.e., the abnormal sample detected by the GRU-based autoencoder) is firstly fed into 2 BiGRU layers which are composed of 64 and 32 cells, respectively, to learn the relevant sequential features $[\bm{h}_1, \bm{h}_2,\ldots,\bm{h}_{31}]$. 
Then, 
in the attention layer,
the extracted feature $\bm{h}_i$ attains  
a weight (i.e., attention score) $\alpha_i$  
which is calculated as follows:
\begin{eqnarray}
\bm{e}_i &=& tanh(\bm{W}_h \bm{h}_i)\\
\alpha_i &=& softmax(\bm{w}^T \bm{e}_i)
\end{eqnarray}
where $\bm{W}_h$ and $\bm{w}$ denote weight matrices, and
$softmax$ represents a normalizing function 
for ensuring that $\alpha_i \geq 0$,  and $\sum_i \alpha_i = 1$. 
By aggregating the computed weights $\bm{\alpha}_i$ and $\bm{h}_i$, 
a weighted feature vector (i.e. attention context vector) $\bm{c}$ is computed as follows:
\begin{equation}
\bm{c} = \sum_i \bm{\alpha}_i \bm{h}_i, 
\end{equation}
which captures the relevant information to improve the performance of the model.

Afterwards, the vector $\bm{c}$  is transferred to two task-specific layers which are dedicated to solving the task $T_1$ and $T_2$ by leveraging the knowledge extracted from the attention-based BiGRU shared layers. 
The model is trained by minimizing the loss function formulated as:
\begin{equation}
L_{total} = \lambda_1 L_{T_1}   + \lambda_2  L_{T_2}, 
\end{equation}
where $L_{T_1}$ and $L_{T_2}$ represent the loss of  $T_1$ and $T_2$. 
The $L_{T_1}$ is the cross-entropy loss whereas the $L_{T_2}$ is the regression loss in MSE. 
The loss weights $\lambda_1$ and $\lambda_2$ are hyperparameters to be tuned.

\section{Experiments}
\subsubsection{Experimental setup}
\begin{figure*}[!t]
\centering
\includegraphics[width=6.5in]{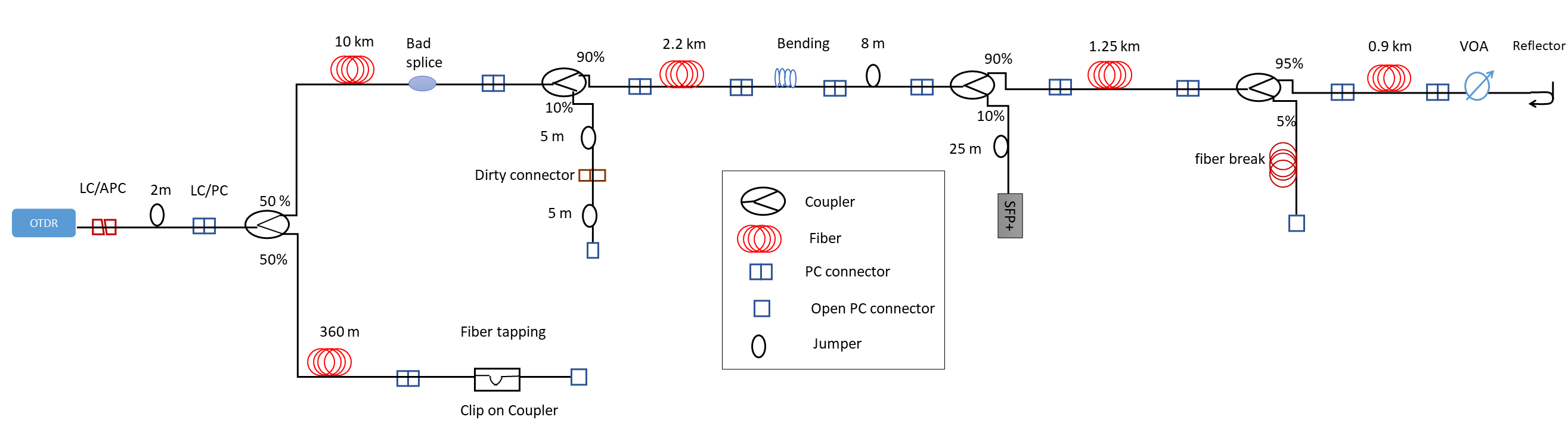}
\caption{Experimental setup for generating OTDR data containing different faults induced at different locations in an optical network (PC - physical contact, APC – angled PC, LC – Lucent connector, SFP+ – small form-factor pluggable)}
\label{fig-4}
\end{figure*}

To validate the proposed approach, an experimental setup is established, as shown in Figure \ref{fig-4}. 
The setup is used to record OTDR traces incorporating different types of fiber faults namely fiber cut, fiber eavesdropping (fiber tapping), dirty connector and bad splice. 
To reproduce a real passive optical network environment, four couplers are employed. 
Optical components such as connectors, a variable optical attenuator (VOA) and a reflector are utilized to model normal events in the fiber optic link.  
To vary the fiber bending pattern and thereby to enhance the generalizability capability of the ML model, the bend radius of the clip-on coupler is ranged from 2.5 mm to 10 mm. 
Different bad splices with dissimilar losses are performed to create a varying bad splicing fault pattern. 
The OTDR configuration parameters, namely the pulse width, the wavelength and the sampling time, are  set  to  10  ns,  1650  nm  and  1  ns, respectively. 
The OTDR records from 62 up to 65,000 are collected and averaged. 

\subsubsection{Data Preprocessing}
The generated OTDR traces are segmented by every 30 traces and normalized to form a sequence.
For each sequence, its SNR $\gamma$ is computed and assigned. 
The GRU-based autoencoder (GRU-AE) is trained with only the normal sequences, that is, a series of traces induced in a normal state of the optical components without fault.
Whereas, for testing, both normal and abnormal sequences are used to incorporate an anomaly. 
For training GRU-AE, 47,904 traces are generated in total and split into a training dataset (70\%) and a test dataset (30\%). 
For training the attention-based BiGRU model, we consider only the faulty sequences. 
Each sequence is assigned with the fault type (fiber eavesdropping, bad splice, fiber cut, dirty connector) and the fault position, which is defined as the index within the sequence.  
To train the fault diagnosis and localize an ML model, total 61,849 traces are used, where those data are divided into a training (60\%), a validation (20\%) and a test dataset (20\%).

\section{Performance Evaluation}

\subsubsection{Evaluation metrics} 
The fault detection is modeled as a binary classification.
The “positive” sequences are labeled with “1: fault”, whereas the “negative” sequences are with “0: normal”.
Then, the following classes are considered for the evaluation metric:
\begin{itemize}
\item True positives (TP): number of sequences of type ‘1’ correctly classified with label ‘1’;
\item True negatives (TN): number of sequences of type ‘0’ correctly classified with label ‘0’;
\item False positives (FP): number of sequences of type ‘0’ misclassified with label ‘1’;
\item False negatives (FN): number of sequences of type ‘1’ misclassified with label ‘0’.
\end{itemize}
To assess the detection capability, the following metrics are adopted.
\begin{description}
\item{\bf Precision (P): } $P$ quantifies the relevance of the predictions made by the ML model. It is expressed as:
\begin{equation}
P = \frac{TP}{(TP +  FP)}
\end{equation}
\item{\bf Recall (R): } $R$ provides the total relevant results correctly classified by the ML model. It is formulated as:
\begin{equation}
R = \frac{TP}{(TP +  FN)}
\end{equation}
\item{\bf F1 score: } $F1$ is the harmonic mean of the precision and recall, calculated as:
\begin{equation}
F1 = 2 \cdot \frac{P\cdot R}{(P + R)}
\end{equation}
\end{description}

\subsubsection{Fault detection capability} 
\begin{figure}[H]
\centering
\includegraphics[width=0.9\columnwidth]{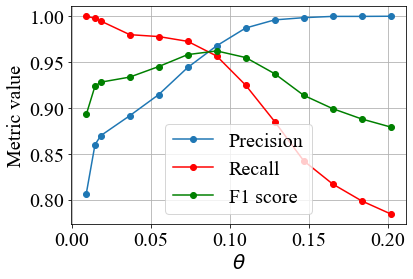}
\caption{The optimal threshold selection based on the precision, recall and F1 scores yielded by GRU-AE.}
\label{fig-6}
\end{figure}

The anomaly detection capability of GRU-AE is varied depending on 
the selected threshold $\theta$. 
Figure \ref{fig-6} shows the precision, the recall, and the F1 score 
curves as function of $\theta$.
If the selected threshold is too low, many faults will be classified as normal, leading to a higher false positive ratio. 
Whereas, if the chosen threshold is too high, many "normal" sequences will be classified as "faulty", resulting in a higher false negative ratio. 
Therefore, the optimal threshold that ensures the best precision and recall tradeoff (i.e., maximizing the F1 score) should be chosen. 
According to Figure \ref{fig-6}, the fault detection capability is optimal when $\theta$ is equal to $0.09$, and the precision, the recall, and the F1 scores are 96.8\%, 95.7\%, and 96.2\%, respectively.

\begin{figure}[H]
\centering
\includegraphics[width=0.9\columnwidth]{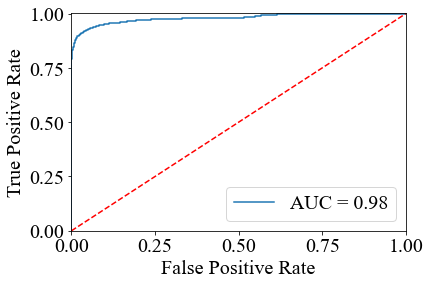}
\caption{The receiver operating characteristic curve of GRU-AE}
\label{fig-7}
\end{figure}

Another way to illustrate the performance of the model is to draw a receiver operating characteristic (ROC) curve  at different threshold settings. 
Figure \ref{fig-7} shows that the distinguishability of GRU-AE between the normal and faulty classes is outstanding
since an area under the curve (AUC) (i.e., degree of separability between classes) is as high as 0.98.

\subsubsection{Fault diagnosis capability} 
\begin{figure}[!t]
\centering
\includegraphics[width=0.9\columnwidth]{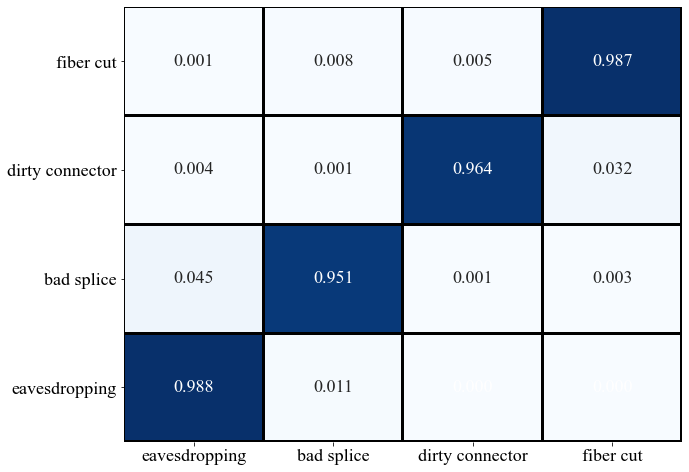}
\caption{The confusion matrix of the attention-based BiGRU model}
\label{fig-9}
\end{figure}

The confusion matrix of the attention-based BiGRU model (A-BiGRU) is illustrated in Figure \ref{fig-9}. 
Our experiments show that the A-BiGRU model identifies the different faults with an accuracy higher than 95\% and 
the fiber physical attacks with an accuracy higher than 98\%. 
As for the low SNR sequences, the ML model mis-classified these classes with 1.1\% probability since there are similarities between the patterns of the eavesdropping and the bad splice faults. 
The same justification applies for the patterns of a dirty connector and a fiber break under low SNR conditions. 

\begin{figure}[H]
\centering
\includegraphics[width=0.9\columnwidth]{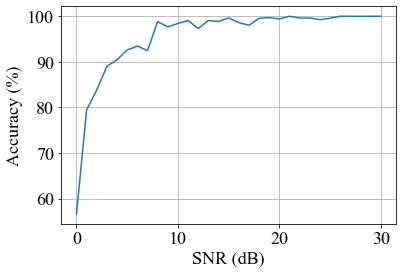}
\caption{Fault diagnosis performance evaluation}
\label{fig-10}
\end{figure}

Figure \ref{fig-10} shows the effects of SNR on the diagnosis accuracy of A-BiGRU. 
The accuracy increases with the SNR. 
For the SNR values higher than 10 dB, the accuracy is approaching to 1. 
For the SNR lower than 2 dB, the accuracy is sharply dropped as it is difficult to differentiate the fault types by the patterns of the faults mainly due to the noise which adversely impacts the patterns under very low SNR levels.

\begin{figure}[!t]
       \begin{subfigure}[b]{0.23\textwidth}
            \centering
            \includegraphics[width=\textwidth]{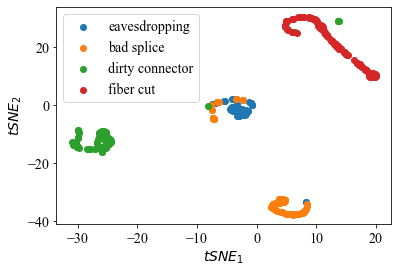}
            \caption[]%
            {{\small 5 dB SNR}}    
        \end{subfigure}
        \hfill
        \begin{subfigure}[b]{0.23\textwidth}  
            \centering 
            \includegraphics[width=\textwidth]{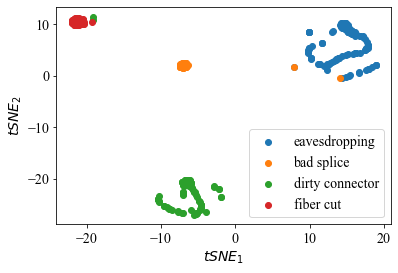}
            \caption[]%
            {{\small 10 dB SNR}}    
        \end{subfigure}
        \vskip\baselineskip
        \begin{subfigure}[b]{0.23\textwidth}   
            \centering 
            \includegraphics[width=\textwidth]{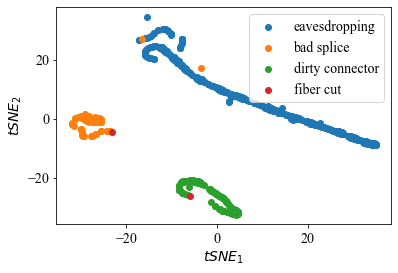}
            \caption[]%
            {{\small 15 dB SNR}}    
        \end{subfigure}
        \hfill
        \begin{subfigure}[b]{0.23\textwidth}   
            \centering 
            \includegraphics[width=\textwidth]{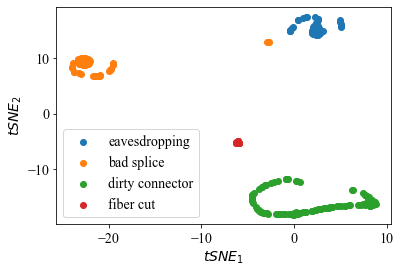}
            \caption[]%
            {{\small  25 dB SNR}}    
        \end{subfigure}
\caption{Visualization of the feature learning under different SNR conditions
}
\label{fig-11}
\end{figure}

The feature learning ability of A-BiGRU for solving the task $T_1$ is investigated  using the t-distributed stochastic neighbor embedding (t-SNE) technique under different SNR conditions \cite{tsne}.
As compared in Figure \ref{fig-11}, 
the learned features become more and more prominent with the increase of the SNR. Furthermore,  A-BiGRU is able to learn effective features for accurate fault diagnosis even under low SNR conditions. 

\subsubsection{Fault localization capability} 
\begin{figure}[H]
\centering
\includegraphics[width=0.9\columnwidth]{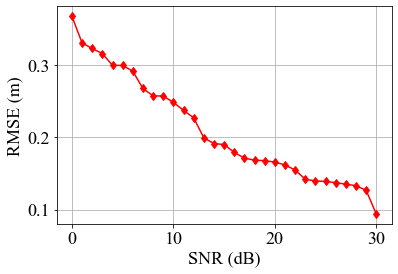}
\caption{Fault position estimation error (in RMSE) for the ML model}
\label{fig-12}
\end{figure}

To measure the accuracy of the fault localization capability of A-BiGRU,
an average root mean square error (RMSE) is introduced.
Figure \ref{fig-12} shows that 
A-BiGRU localizes the faults accurately by achieving the RMSE of 0.21m. 
For lower SNR values (e.g., SNR $\le$ 10 dB), the RMSE can be higher than 0.37m, whereas for SNR values higher than 15 dB, it is less than 0.2m, and it could be further reduced up to less than 0.1m for SNR values higher than 30 dB.

\section{Conclusion}
Fiber monitoring is essential for long-lasting optical network operations and high service availability. 
Our experiments show that ML techniques can enhance the performance of the anomaly detection and fault identification in fiber monitoring, and minimizing the false positive alarm.
For the future work, we apply our framework and methods for the  cybersecurity of a cyber-physical system (CPS) which is usually operated in a very complex, sophisticated, intelligent and autonomous environment.

\bibliography{bc}

\section{Acknowledgments}
This work has been performed in the framework of the CELTIC-NEXT project AI-NET-PROTECT (Project ID C2019/3-4), and it is partly funded by the German Federal Ministry of Education and Research (FKZ16KIS1279K).
\end{document}